\documentclass[aps,showpacs,nofootinbib]{revtex4}%
\usepackage{graphicx}
\usepackage{amsmath}
\usepackage{amssymb}
\usepackage{amsfonts}%
\newcommand{\be}{\begin{equation}}
\newcommand{\ee}{\end{equation}}
\newcommand{\ba}{\begin{eqnarray}}
\newcommand{\ea}{\end{eqnarray}}

\begin{document}
\title{Signal transmission on lossy lines as a dissipative quantum state propagation.}
\author{Yu. Reznykov.}
\email{yuriy@physik.tu-cottbus.de}
\affiliation{Brandenburgische Technische Universit\"{a}t, Cottbus, Germany.}
\date{\today}

\begin{abstract}
The transmission of electric signals on a coupled line with distributed
RLC-parameters is considered as a propagation of a dissipative quasi particle.
A calculation technique is developed, alternative to the one, accepted for
lumped lines. The relativistic wave equation for the transient response is
deduced following the common Ohm-low-type considerations. The exact
expressions for the Green function, for information transfer velocity and for
time delay are obtained on this base. The fundamental restrictions on the
measurement accuracy of the time delay are pointed out. The obtained results
are naturally generalized for the multilevel networks of the arbitrary dimension.

\end{abstract}

\maketitle

\section{Introduction}

The electric signals transmission on distributed RLC-lines will be solved
analytically. Our goals are deformation of a signal shape and time delay.
These questions are usually considered by the calculation technique, made for
lumped lines [1-5]. On the other hand, in the works [6-9] people build models,
based on the Maxwell- or Schr\"{o}dinger-like equations.

The physical approach, developed here, is based on the introduction of
electric impulses on the lossy line as a dissipative quasi particle. The
space-time propagation of it is described by the Klein-Gordon equation - the
equation of motion of free relativistic scalars (see, for example, [10]). The
parameters of an electric signal have a direct particle-like meaning. For
example, the time decay rate plays the role of a negative mass square.

\section{Equations, solutions.\ Information transfer}

The lossy interconnect line can be introduced as consisting of a row of
infinitesimally small RLC-cells.\ Each cell of length $dl$\ has a local
resistance $dR,$ inductance $dL$ and capacitance $dC:$%

\begin{equation}
{\normalsize dR}=r(l)dl,\qquad{\normalsize dL=}\pounds (l)dl,\qquad
{\normalsize dC=}c(l)dl,
\end{equation}
determined by the linear densities as:%

\begin{equation}
r(l)=dR/dl,\qquad\pounds (l)=dL/dl,\qquad c(l)=dC/dl.
\end{equation}
The target voltage of cell $U_{C}=$ $U_{out}(t)$\ is expressed through
entrance one - $U_{in}(t)$ by the condition:%
\begin{equation}
{\normalsize U}_{in}{\normalsize =U}_{L}+{\normalsize U_{R}}+{\normalsize U}%
_{out},
\end{equation}
where:%
\begin{equation}
{\normalsize U}_{L}{\normalsize =dL}\frac{{\normalsize \partial J}%
}{{\normalsize \partial t}}{\normalsize ,\qquad U_{R}=dRJ,}\qquad
\frac{{\normalsize \partial J}}{\partial l}={\normalsize c\frac{\partial
U}{\partial t}.}%
\end{equation}
So, the voltage $U(l,t),$ as a function of time and distance, obeys the
equation:%
\begin{equation}
{\normalsize c}\pounds \frac{{\normalsize \partial}^{2}{\normalsize U}%
}{{\normalsize \partial t}^{2}}{\normalsize -}\frac{{\normalsize \partial}%
^{2}{\normalsize U}}{\partial l^{2}}{\normalsize +rc}\frac
{{\normalsize \partial U}}{{\normalsize \partial t}}=0.
\end{equation}

This equation is known as the one-dimensional telegraph equation and describes
the wave distribution with damping. For the case $\pounds =0$ it has been
treated in [1]. The space variable can be normalized by replacing $l$ with
$x=l/v$. Then, using the notations: $m=r/2\pounds $ $,v^{2}=1/c\pounds $, the
function $\Phi(x,t)$ is introduced: $U(x,t)=\exp(-mt)\Phi(x,t).$ It obeys the equation:%

\begin{equation}
\frac{{\normalsize \partial}^{2}{\normalsize \Phi}}{{\normalsize \partial
t}^{2}}-\frac{{\normalsize \partial}^{2}{\normalsize \Phi}}%
{{\normalsize \partial x}^{2}}-{\normalsize m}^{2}{\normalsize \Phi=0.}%
\end{equation}

This is the Klein-Gordon equation* with the negative mass square - so-called
"tachyons". Earlier we dealt with such an exotic object in a real situation
[10]. This representation looks like the motion of the tachyon quasi particle
in the constant external dissipative field.

The basis solution of the main equation (5) with the wave vector $k$%

\begin{equation}
\varphi({\normalsize \omega}_{{\small 0}}{\small \mid}{\normalsize {\small x}%
,}{\small t)}={\normalsize e}^{-mt{\normalsize -i(\omega}_{{\small 0}%
}{\normalsize t-{\small k}x)}},\text{ \ \ }{\normalsize \omega}_{0}%
(k)=\sqrt{{\small k}^{2}-{\small m}^{2}}%
\end{equation}

is a plane wave, damped in time at the rate $m$, identical for all
frequencies. A macro-packet - $U(x,t)$ can be built as a linear superposition
of basis states:%
\begin{equation}
{\normalsize U(}x{\normalsize ,{\small t})=}\underset{\underline
{{\normalsize \omega}}}{\overset{\overline{{\normalsize \omega}}}{\int}%
}d{\normalsize \omega}_{{\small 0}}{\normalsize U(\omega}_{{\small 0}%
}{\normalsize )}\varphi({\normalsize \omega}_{{\small 0}}{\small \mid
}{\normalsize {\small x},}{\small t)}.
\end{equation}
The wave $\varphi(\omega_{{\small 0}}\mid x,t)$ is a not a stationary state,
but its energy loss in time as the square of amplitude:%
\begin{equation}
{\normalsize \hat{\omega}}\varphi({\normalsize \omega}_{{\small 0}%
}{\small \mid}{\normalsize {\small x},}{\small t)}={\normalsize \omega
}(t)\varphi({\normalsize \omega}_{{\small 0}}{\small \mid}%
{\normalsize {\small x},}{\small t)}={\normalsize e}^{-{\small 2mt}%
}{\normalsize \omega}_{{\small 0}}\varphi({\normalsize \omega}_{{\small 0}%
}{\small \mid}{\normalsize {\small x},}{\small t)}.
\end{equation}
\ The frequency-width of a packet: $\Delta\omega=\overline{{\normalsize \omega
}}-\underline{{\normalsize \omega}}-$ also falls down in time as
$e^{-{\small 2mt}}.$ The packet does not spread in time, but agglomerates in
frequencies. The wave $\varphi(\omega_{{\small 0}}\mid x,t)$ has a phase
velocity:
\begin{equation}
{\normalsize v}_{ph}={\normalsize \omega}_{{\small 0}}{\normalsize /k=\omega
}_{{\small 0}}{\normalsize /}\sqrt{{\normalsize \omega}_{{\small 0}}%
^{2}+{\small m}^{2}}.
\end{equation}
But the information transfer velocity- the group velocity of a packet, defined
as: $u\equiv\partial\omega_{{\small 0}}/\partial k,$ for the dispersion law
(7) results:%
\begin{equation}
{\normalsize v}_{gr}{\normalsize =}\sqrt{{\normalsize \omega}_{{\small 0}}%
^{2}+{\small m}^{2}}{\LARGE /}{\normalsize \omega}_{{\small 0}}%
=k{\normalsize /\omega}_{{\small 0}}{\normalsize .}%
\end{equation}
On the first view this velocity (in the natural units: $v_{gr}=vk/\omega
_{{\small 0}}$) is higher than the light speed in the wire: $v_{gr}>v$. But
the packet decays in time as $e^{-mt}$. In a short time $dt$ it loses the
part: $dx_{2}\approx\frac{\pi m}{{\normalsize \omega}_{{\small 0}}}dt$ from
the forward edge, which amplitude becomes $e^{-1}$ of initial. This part must
be subtracted from the whole packet relocation: $dx_{1}=v_{gr}dt$. So, the
effective speed of the packet can be approximated as:%

\begin{equation}
u\approx{\normalsize v}_{gr}-\frac{\pi m}{{\normalsize \omega}_{{\small 0}}%
}=(\sqrt{{\normalsize \omega}_{{\small 0}}^{2}+{\small m}^{2}}-{\small \pi
}m){\LARGE /}{\normalsize \omega}_{{\small 0}}<1.
\end{equation}
\qquad

The treatment of electric impulses on the lossy line as a quasi particle gives
a physical restriction on the measurement accuracy of the time delay.
Following the uncertainty relations, the measurement of the time-slice $\Delta t$ is
accompanied with the energy variation: $\Delta\omega_{{\small 0}}%
\gtrsim1/\Delta t.$ At the signal frequency $\omega_{{\small 0}}$ the delay
time $\delta$ can be measured up to accuracy:$\Delta\delta$ $\gtrsim
1/\omega_{{\small 0}}.$ This inequality gives a minimal error, physically
permissible by the measurements of time-slices. This restriction has a
fundamental character, and does not depend on the chosen approach.

\section{Green-function. Integral representation}

The Green function for the equation (5) can be built from the solutions (7) as:%

\begin{equation}
{\normalsize G(x,t)=}\frac{{\normalsize e}^{-mt}}{({\normalsize 2\pi
}{\large )}^{2}}\underset{-\infty}{\overset{\infty}{\int}}{\normalsize d\omega
}\underset{-\infty}{\overset{\infty}{\int}}{\normalsize dk}\frac
{{\normalsize \exp[-i(\omega t-{\small k}x)]}}{{\normalsize -}{\small m}%
^{2}{\normalsize -\omega}^{2}{\normalsize +}{\small k}^{2}}.
\end{equation}
The integrand has two poles at $\omega=\omega_{1/2}$, which form two complex
resonance frequencies:
\begin{equation}
{\normalsize \omega}_{1/2}(k)=-mi\pm{\normalsize \omega}_{0}(k).
\end{equation}
At $k<m$ frequency ${\normalsize \omega}_{0}$ becomes imaginary- the wave does
not propagate. But for all $-\infty<k<\infty$ the both poles lay below the
real $\omega$-axis. To build a retarded Green function, the integration
contour can be kept on the real $\omega$-axis. At $t>0$ the contour closes in
the low $\omega$-half-plane and contains all possible positions of poles. The
retarded Green function is the sum of the residues in the both poles
$\omega_{1/2}$:
\begin{equation}
{\normalsize G}^{ret}{\normalsize (x,t)}=\frac{{\normalsize e}^{-mt}%
{\small \Theta(t)}}{{\small 2}{\normalsize \pi}}\underset{-\infty}%
{\overset{\infty}{\int}}{\normalsize d}{\small k}\frac{\sin{\normalsize \omega
}_{0}{\small t}}{{\normalsize \omega}_{0}}e^{ikx},
\end{equation}
where: $\Theta(t)$- is a Heaviside function. The real range of the integration
in (15) is restricted by the condition: $|k|\geq m$. Such "damped tachyon"
Green function in the coordinate representation is the analytical continuation
of the scalar one into the area $m^{2}\rightarrow-m^{2}$. Using the integrals,
calculated in [10], this function can be expressed as:%

\begin{equation}
{\normalsize G}^{ret}{\normalsize (x,t)}=-{\small \Theta(t)\Theta(\lambda
)}{\normalsize e}^{-mt}I_{0}(m\sqrt{{\small \lambda}}){\small ,}%
\end{equation}
where: $\lambda=t^{2}-x^{2},$ and $I_{0}(x)-$ modified Bessel - function.

Supposing that the source voltage $U_{in}(t)$ was applied to the RLC-line
during the limited time $0\leqslant t\leqslant t_{ac}$\ at the point $x=0:$%
\begin{equation}
{\normalsize U}_{in}{\normalsize (x=0,t)=}u_{0}{\normalsize (t).}%
\end{equation}
The cross-section of the wire in the point $x=0$ can be treated as a
2-dimensional boundary $(S)$.\ The unit vector of an external normal to it is
denoted as $\overrightarrow{n}.$ The solution of homogeneous equations (5) at
the boundary condition (17) is given by the integral formula:%
\begin{equation}
{\normalsize U(x,t)=}\underset{0}{\overset{{\normalsize \tau}}{\int}%
}{\normalsize dt%
%TCIMACRO{\U{b4}}%
%BeginExpansion
\acute{}%
%EndExpansion
G}^{S}{\normalsize (x,t%
%TCIMACRO{\U{b4}}%
%BeginExpansion
\acute{}%
%EndExpansion
)u_{0}(t-t%
%TCIMACRO{\U{b4}}%
%BeginExpansion
\acute{}%
%EndExpansion
),}\text{ \ \ }{\normalsize \tau=\max(t,t}_{ac}),
\end{equation}
where $G^{S}(x,t)$ -is the Green function of the 2nd kind, expressed through
$G^{ret}$ as: $G^{S}(x,t)\equiv-\frac{{\normalsize dG}^{ret}}{dn}.$ Taking
into account the property of Bessel - functions and noting, that
$\overrightarrow{n}-$ indicates the positive $x$-direction, it becomes:%
\begin{equation}
{\normalsize G}^{S}{\normalsize (x,t)}={\normalsize 2{\small x}e^{-m{\small t}%
}[\delta}{\small (\lambda)+\Theta(\lambda)}I_{1}(m\sqrt{{\small \lambda}})].
\end{equation}

\section{Delays. $\ $Leading transfer mode}

The representation (19) can be applied to the macro-packet (8). The amplitude
of a signal, propagating in the positive (negative-) $x$-direction-
$U_{{\normalsize r}}(U_{{\small l}})$, is equal:%
\begin{equation}
{\normalsize U}_{r(l)}({\normalsize x,t})={\normalsize e}^{-mx}%
{\normalsize u_{0}}{\small (t\mp x)+x}\underset{{\small 0}}{\overset
{{\small \tau}^{2}{\small -x}^{2}}{\int}}d\lambda\frac{{\normalsize e}%
^{-m\sqrt{\lambda+{\small x^{2}}}}}{{\small \sqrt{\lambda+x^{2}}}}\ast
\end{equation}%
\[
\ast I_{1}(m\sqrt{\lambda}){\normalsize u_{0}}{\small (t-\sqrt{\lambda+x^{2}%
})}{\normalsize ,}\text{ }x<\tau.
\]
Such a representation repeats the result, given in [3]. The input signal
$U_{in}$ in the point $x$ achieves the level $U_{in}=bU_{\max}$, $b<1$ in the
delay time $\delta$, determined by the equation:%
\begin{equation}
U_{{\normalsize r}}({\normalsize x,}\delta)={\normalsize bU}_{\max}.
\end{equation}

The frequent properties of the transient response, defining by the Green
function (15), are following. The signal of frequency $\omega_{in}$ is
transmitted by the leading transfer mode (LTM) of the same frequency:
$\omega_{0}=\omega_{in}.$ For example, at small frequencies: $\omega
_{in}\longrightarrow0,{\normalsize u_{0}=const,}$ the asymptotic of the
Green-function, also corresponding to the DC propagation, has a view:%

\begin{equation}
{\normalsize G}_{0}^{ret}{\normalsize (x,t)\approx-}\frac{{\small t}%
{\normalsize e}^{-mt}}{{\normalsize \pi}}\frac{\sin{\small m}x}{x},\text{
}{\normalsize \omega}_{0}\longrightarrow0{\normalsize .}%
\end{equation}
For the constant input signal ${\normalsize u_{0}}$ equation for the delay
time in the point $x$\ is:%
\begin{equation}
\frac{{\small t}{\normalsize e}^{-mt}}{{\normalsize \pi}}[\frac{\sin
{\small m}x}{x^{2}}-\frac{m\cos{\small m}x}{x}]{\LARGE [}({\small t+1/m)}%
{\normalsize e}^{-mt}-{\small 1/m}{\LARGE ]}=b
\end{equation}

\section{Reflections}

For the finite RLC-lines, length $\bar{l}$ $($ $\bar{x}=l/v),$ reflections
occur. The wave packet is now a superposition of incoming and reflected waves.
The resulting voltage with $N_{r}$- reflections is:%

\begin{equation}
{\normalsize U}_{ref}({\normalsize x,t})=\overset{N_{r}}{\underset
{{\small s=0}}{%
%TCIMACRO{\tsum }%
%BeginExpansion
{\textstyle\sum}
%EndExpansion
}}{\LARGE [}\Gamma^{2{\normalsize s}}U_{{\normalsize r}}({\small x+2s\bar
{x},t})+\Gamma^{2{\normalsize s+1}}U_{l}({\small 2s\bar{x}-x,t}){\LARGE ]}%
{\normalsize ,}%
\end{equation}
where $\Gamma-$ is the reflection coefficient. The flying signal decays on the
length: $\bar{l}_{d}\sim v/m$. For the line of length $\bar{l},$ the
reflection number is the ratio: $N_{r}\approx\bar{l}_{d}/\bar{l}.$ Equation:
$U_{{\normalsize ref}}(x,\delta)=bU_{\max}$ gives now the signal delay time
$\delta.$

In the modern interconnection RLC-lines, for example,\ at the parameters
values as in [4,5]: $r=37.8$ $\Omega/cm,$ $c=3.28e-13F/cm,$ $Z_{0}%
=\sqrt{\pounds /c}=266.5\Omega$ and the length of the line: $\overset{\_}%
{l}=3.6cm, $ we have: $\pounds =2.3e-8Hn/cm,$ the decay rate:$\ m=8.1e+08Hz$.
For the signal frequency $\nu\approx3GHz$\ the decay length is: $\bar{l}%
_{d}\approx14.1cm$ - and, at least: $N_{r}\approx4$ reflections are
observable$.$

\section{Multilevel networks}

For generalizing the obtained results on the multilevel network, the network
can be represented as the $N$-component space formed from $N$ $RLC$-lines. An
orthogonal basis will be put on the directions of RLC-lines:
$\{e\}=\{\overrightarrow{e}_{1},\overrightarrow{e}_{2}...\overrightarrow
{e}_{N}\},$ $\overrightarrow{e}_{i}\ast\overrightarrow{e}_{k}=\delta^{ik}$.
The voltage impulse, flowing through the network, becomes now a vector in this
space. For example, for the network, consisting of three lines, the voltage
vector is: $\overrightarrow{{\small U}}(l,t)=[U_{1}(l,t),U_{2}(l,t),U_{3}%
(l,t)].$ The vector $\overrightarrow{{\small U}}$ obeys the equation (5) in
the matrix form [5]:%

\begin{equation}
\widehat{\mathbf{c}}\widehat{\mathbf{\pounds }}\frac{{\normalsize \partial
}^{2}\overrightarrow{{\small U}}}{{\normalsize \partial t}^{2}}{\normalsize -}%
\frac{{\normalsize \partial}^{2}\overrightarrow{{\small U}}}{\partial l^{2}%
}{\normalsize +r}\widehat{\mathbf{c}}\frac{{\normalsize \partial
}\overrightarrow{{\small U}}}{{\normalsize \partial t}}=0,
\end{equation}
where the matrixes $\widehat{{\normalsize c}}$ and $\widehat{\pounds },$ are:%
\begin{equation}
\widehat{\mathbf{c}}{\small =}\left[
\begin{array}
[c]{ccc}%
{\small 2c}_{grd}{\small +c}_{m} & {\small -c}_{m} & {\small 0}\\
{\small -c}_{m} & {\small 2c}_{grd}{\small +2c}_{m} & -c_{m}\\
{\small 0} & {\small -c}_{m} & {\small 2c}_{grd}{\small +c}_{m}%
\end{array}
\right]  {\small ,}%
\end{equation}%
\begin{equation}
\widehat{\mathbf{\pounds }}{\small =}\left[
\begin{array}
[c]{ccc}%
\pounds _{{\small 11}} & \pounds _{{\small 12}} & \pounds _{{\small 13}}\\
\pounds _{{\small 12}} & \pounds _{{\small 22}} & \pounds _{{\small 23}}\\
\pounds _{{\small 13}} & \pounds _{{\small 23}} & \pounds _{{\small 33}}%
\end{array}
\right]  ,
\end{equation}
$c_{grd}$ - line-to-ground capacitance, $c_{m}$- line-to-line capacitance; and
$\pounds _{{\small ik}}$-self- and mutual between-conductor-inductances. The
non-diagonal components of tensors describe interline transmission. It was
pointed out in [5], the product of the matrixes $\widehat{\mathbf{c}}$ and
$\widehat{\mathbf{\pounds }}$ is proportional to the unit matrix:
$\widehat{\mathbf{c}}\widehat{\mathbf{\pounds }}=\frac{1}{{\normalsize v}^{2}%
}\widehat{\mathbf{I}}.$. It is natural as voltage obeys the inhomogeneous
Maxwell equation, containing, as well as (6), D`Alamber operator. Then the
space variable $l$ can again be normalized by: $x=l/v$ and the particle mass
(which is now a tensor) can be introduced: $\widehat{\mathbf{m}}%
=\frac{r{\normalsize v}^{2}}{2}\widehat{\mathbf{c}}=\frac{r}{2}\widehat
{\mathbf{\pounds }}^{-1}.$ The multilevel network with distributed
RLC-parameters appears as an anisotropic solid medium. The situation with the
quasi particle parameters, depending on direction, is here well known. By
substitution: $\overrightarrow{{\small U}}(x,t)=$ $\exp(-\widehat{\mathbf{m}%
}t)\overrightarrow{{\small \Phi}}(x,t)$ the matrix analog of the equation (6)
for $\overrightarrow{{\small \Phi}}$ can be obtained:%

\begin{equation}
\frac{{\normalsize \partial}^{2}\overrightarrow{{\small \Phi}}}%
{{\normalsize \partial t}^{2}}-\frac{{\normalsize \partial}^{2}\overrightarrow
{{\small \Phi}}}{{\normalsize \partial x}^{2}}-\widehat{\mathbf{m}}%
^{2}\overrightarrow{{\small \Phi}}{\normalsize =0.}%
\end{equation}

The vectors $\overrightarrow{{\small U}}(x,t)$ and $\overrightarrow
{{\small \Phi}}(x,t)$ in general case are not parallel, so as the decay
operator $\exp(-\widehat{\mathbf{m}}t)$ mix's the components. It seems that
the last term in (28) describes the self-interaction of the field
$\overrightarrow{{\small \Phi}}.$ But the numerical tensor $\widehat
{\mathbf{m}}$, generally, can be diagonalized and thus the field components
will be split. With it the Green function will also be diagonalized. Then, the
proper directions of the tensor $\widehat{\mathbf{m}}$: $\{\epsilon
\}=\{\overrightarrow{\epsilon}_{{\small 1}},\overrightarrow{\epsilon
}_{{\small 2}},\overrightarrow{\epsilon}_{{\small 3}}\}$ can be chosen as the
new basis in the network-space. The same quasi-free propagation occurs along
the each of new orts as in the case of the single RLC-line:
\begin{equation}
{\normalsize U}_{i}{\normalsize (x,t)=}\exp{\normalsize (-}{\small m}%
_{i}{\normalsize t)\Phi}_{i}{\normalsize (x,t).}%
\end{equation}
The decay rates $(m_{1},m_{2},m_{3})$ are different for each ort of this
basis. Now the further reasons and results of the work can be reproduced for
multilevel networks. The Green function has the same form as (15):%
\begin{equation}
\overrightarrow{{\normalsize U}}_{(m)}({\normalsize x,t})=\underset
{0}{\overset{{\normalsize \tau}}{\int}}dt%
%TCIMACRO{\U{b4}}%
%BeginExpansion
\acute{}%
%EndExpansion
{\normalsize G}^{ret}{\normalsize ({\small m}_{i}\mid x,t-t%
%TCIMACRO{\U{b4}}%
%BeginExpansion
\acute{}%
%EndExpansion
)}\overrightarrow{{\normalsize U}}_{(in)}({\normalsize t%
%TCIMACRO{\U{b4}}%
%BeginExpansion
\acute{}%
%EndExpansion
}),
\end{equation}
where $\overrightarrow{{\small U}}_{(in)}(t)$- is incoming voltage vector, in
the new basis $\{\epsilon\}$. The traveling voltage, propagating in the real
space on the RLC-lines, becomes the projection of $\overrightarrow{{\small U}%
}_{(m)},$ calculated by (30), on the corresponding unit vector of the old basis
$\{e\}$.

\section{Conclusion}

The suggested representation for the transient response of the distributed
RLC-line has allowed to obtain the exact formulas for the Green function and
time delay. The simple description of signals propagation by the frequency
characteristics in the LTM-approximation is obtained. The natural
generalization on the multilevel networks of arbitrary dimension is made. The
physical approach can be helpful at the multilevel networks design and in the
quantum information.

\begin{acknowledgments}
The author expresses gratitude to professor D. Robaschik for useful discussions.
\end{acknowledgments}

---------------------------------------------

*Indeed, it is necessary to multiple the equation by factor $\hbar^{2}$%
/$c^{2},$ but in the field theory the system of unit is adopted, where:
$\hbar$=1, $c=1.$

\end{document}